\documentclass[sigconf, 9pt]{acmart}

\usepackage{amsmath,amssymb,amsfonts}
\usepackage{algorithmic}
\usepackage{graphicx}
\usepackage{textcomp}
\usepackage{xcolor}
\usepackage{bbm}
\usepackage{multirow}
\usepackage{booktabs}
\usepackage{amsthm}
\usepackage{graphicx}
\usepackage{subcaption}

\newcommand{\yexact}[0]{F_{\textit{exact}}}
\newcommand{\yapprox}[0]{F_{\textit{approx}}}
\newcommand{\ymax}[0]{F_{\textit{MAX}}}

\copyrightyear{2026}
\acmYear{2026}
\setcopyright{cc}
\setcctype{by}
\acmConference[ICCAD '26]{IEEE/ACM International Conference on Computer-Aided Design}{November 08--12, 2026}{San Jose, CA, USA}
\acmBooktitle{IEEE/ACM International Conference on Computer-Aided Design (ICCAD '26), November 08--12, 2026, San Jose, CA, USA}
\acmDOI{10.1145/3831252.3834065}
\acmISBN{979-8-4007-2873-0/2026/11}

\title{Integrating Approximate Logic Synthesis into Approximate High-Level Synthesis}

\author{\large Jian Shi$^{1}$, 
Ruicheng Dai$^{1}$,
Chang Meng$^{2}$,
Yue Yang$^{1}$,
and Weikang Qian$^{1}$}

\affiliation{
\institution{$^{1}$Global College, Shanghai Jiao Tong University, Shanghai, China}
\city{}
\country{}
\institution{$^{2}$Department of Mathematics and Computer Science, Eindhoven University of Technology, Eindhoven, the Netherlands}
\city{}\country{}
}
\email{timeshi@sjtu.edu.cn, ruichengdai@sjtu.edu.cn, c.meng@tue.nl, yue.yang@sjtu.edu.cn, qianwk@sjtu.edu.cn}

\thanks{This work is supported by the National Natural Science Foundation of China under Grant 62574132 and the Natural Science Foundation of Shanghai under Grant 25ZR1401189. Corresponding author: Weikang Qian.}

\begin{document}

\begin{abstract}

Approximate high-level synthesis (HLS) and approximate logic synthesis (ALS) are two techniques for generating approximate circuits.
They operate at different granularities.
Approximate HLS typically modifies instructions in a control and data flow graph, whereas ALS modifies gates and interconnects in a gate-level netlist.
The absence of a unified framework combining these techniques limits the potential for joint optimization.
To bridge this gap, we propose to integrate ALS into the flow of approximate HLS.
This integration expands the design space of approximate HLS by introducing fine-grained approximation induced by ALS, thereby generating approximate circuits with higher quality.
Experimental results show that under the same error bound, our method reduces the hardware cost by 11\% on average compared to the state-of-the-art methods.

\end{abstract}


\maketitle

\section{Introduction}
Many modern applications, such as image processing and machine learning, often require substantial computational resources~\cite{Ahmadinejad22s}.
Fortunately, many of these applications can tolerate certain levels of errors.
\emph{Approximate computing} takes advantage of this error tolerance to trade computational accuracy for improved performance or energy efficiency.

An important area in approximate computing is designing \emph{approximate circuits}.
One approach to this is \emph{approximate logic synthesis} (\emph{ALS}), which automatically generates gate-level approximate circuits with minimized circuit area or delay within a given error bound~\cite{Scarabottolo20s}.
ALS focuses on optimizing gate-level netlists by modifying their internal signals~\cite{SASIMI13s,Chandrasekharan16s,Schlachter17s,ALSRAC20s,ResubALS25s,AccALS25s,Dai25s}.
It explores a large design space, enabling the generation of high-quality approximate circuits, but the vast design space also limits its application to large circuits~\cite{AccALS25s}.

To handle large-scale designs, \emph{approximate high-level synthesis} (HLS) has been proposed~\cite{Schafer26s}.
It is built upon traditional HLS, a technique to automatically convert designs specified in high-level languages (\textit{e.g.}, C/C++) into \emph{register-transfer level} (\emph{RTL}) implementations.
A key step in an HLS flow is transforming high-level languages into a \emph{control and data flow graph} (\emph{CDFG}), which captures the control and data dependencies within the design~\cite{CDFG04s}.
Fig.~\ref{fig:partition}(a) shows an example of a CDFG, where nodes represent different instructions or IO ports and edges represent their dependencies.
Existing approximate HLS methods either apply coarse-grained approximation methods such as precision scaling and variable-to-variable substitution to the CDFG or rely on pre-designed approximate arithmetic units from a library~\cite{Lee06s,Li15s,Lee17s,AxHLS20s,V2VC21s,PreDAC25s,ADVISOR25s}.
This restricted set of approximation choices causes the generated approximate circuits frequently exhibiting suboptimal quality, characterized by limited area reduction and unacceptably high output errors.

To solve the above issues, in this work, we propose to integrate ALS into the flow of approximate HLS and develop a novel approximate HLS method, dubbed \emph{HALS}.
The rationale behind this is that ALS can introduce finer-grained approximation.
Hence, this integration expands the design space of approximate HLS, thereby enabling the generation of approximate circuits with higher quality.
Our main contributions are as follows:
\begin{itemize}
\item We propose a partitioning method tailored for ALS within HLS.
This method partitions a given CDFG into ALS-\linebreak manageable sub-graphs.
\item We introduce a fast error estimation method to evaluate the error impact of ALS-generated circuits within the HLS flow.
\item We develop an efficient \emph{design space exploration} (\emph{DSE}) method to select an optimized combination of ALS-generated candidates during the approximate HLS process.
\end{itemize}

The experimental results show that under the same error bound, our method reduces the hardware cost by 11\% on average compared to the previous state-of-the-art works.
The code of HALS is available at \url{https://www.github.com/SJTU-ECTL/HALS}.

\section{Background and Related Works}\label{sec:review}

\subsection{Approximate High-Level Synthesis}\label{subsec:AxHLS}
\begin{figure}
\centering
\includegraphics[width=\linewidth]{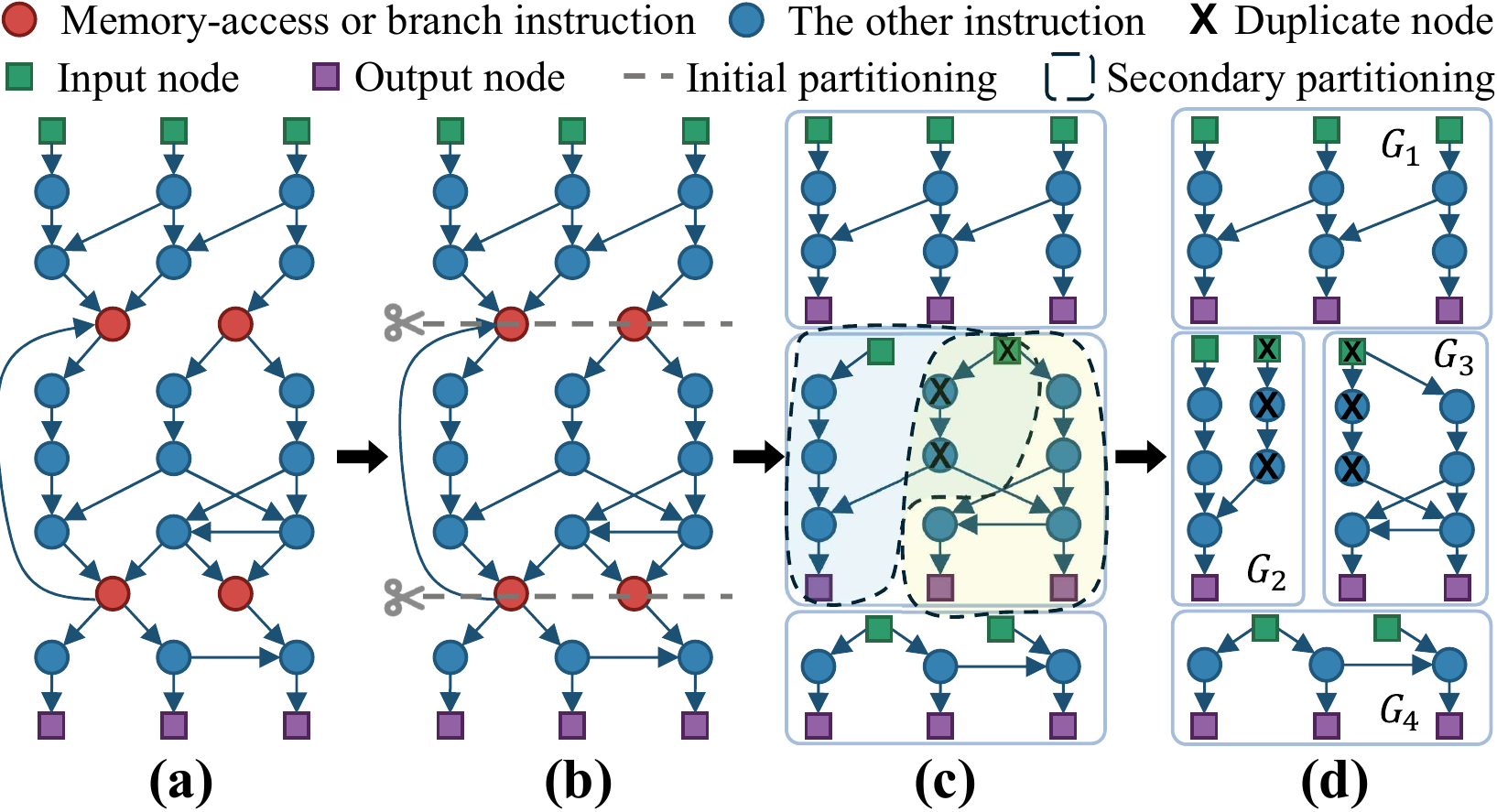}
\vspace{-0.8cm}
\caption{Overview of the graph partitioning in HALS.
The process begins with \textbf{(a)} a CDFG containing nodes and edges representing IR instructions and dependencies, respectively.
\textbf{(b)} The initial partitioning uses memory-access and branch instructions as boundaries.
\textbf{(c)} The secondary partitioning applies the spectral clustering on outputs to further partition large sub-graphs.
\textbf{(d)} The set of partitioned sub-graphs is prepared for the subsequent ALS flow.
}
\vspace{-0.7cm}
\Description{Overview of the graph partition in HALS.}
\label{fig:partition}
\end{figure}

As introduced earlier, HLS transforms high-level specifications (\textit{e.g.}, in C/C++) into RTL designs.
In an HLS flow, it first compiles the source code into a compiler-level \emph{intermediate representation} (\emph{IR}).
Then, it analyzes the control and data dependencies among the IR instructions to construct a CDFG~\cite{Schafer20s}.
As shown in Fig.~\ref{fig:partition}(a), the IR instructions are represented as circles connected by directed edges, which denote their dependencies.
The graph is bounded by input and output ports represented by green and purple squares, respectively.
These ports are essential for modeling but do not consume actual hardware resources.
Once optimized, the CDFG is scheduled and mapped to the final RTL design.

To further improve hardware performance, approximate HLS is proposed~\cite{Schafer26s}.
It directly generates approximate designs from the given high-level specifications.
Existing approximate HLS methods primarily focus on coarse-grained approximations applied to individual nodes or edges within a CDFG.
Major techniques include \emph{precision scaling}~\cite{Lee06s,Li15s}, \emph{mapping instructions to pre-designed approximate function units}~\cite{PreDAC25s,ADVISOR25s,AxHLS20s}, \emph{variable-to-variable/constant substitution}~\cite{ADVISOR25s,V2VC21s}, and \emph{loop perforation}~\cite{LoopPerforation11s,Lee18s,Li18s}.
However, these techniques are applied to nodes in the CDFG individually.
They neglect the substantial design space of joint optimization across multiple consecutive nodes, resulting in suboptimal quality.
To unlock joint optimization, fine-grained gate-level techniques must be exploited in the HLS flow.
This necessitates the integration of ALS.

\subsection{Approximate Logic Synthesis}\label{subsec:ALS}
ALS is an automated technique that optimizes gate-level netlists to minimize hardware costs (\textit{i.e.}, area, delay, and power) while strictly adhering to a given error bound~\cite{Scarabottolo20s}.
Unlike coarse-grained HLS approximations, ALS achieves this by exploring a massive design space through \emph{local approximate changes} (\emph{LACs}), such as replacing a signal with a constant value (0 or 1) or a similar existing signal~\cite{SASIMI13s,Chandrasekharan16s,Schlachter17s,ALSRAC20s,AccALS25s,ResubALS25s}.
These LACs modify local subcircuits, thereby enabling the fine-grained, joint optimization across multiple IR nodes that cannot be achieved by conventional approximate HLS.

To navigate the vast design space, most ALS methods operate in an iterative manner~\cite{SASIMI13s,Chandrasekharan16s,Schlachter17s,AccALS25s,ALSRAC20s,ResubALS25s}.
In each iteration, the ALS methods identify and evaluate a massive number of candidate LACs before applying the modification.
Because this evaluation process is invoked frequently and must capture the complex impact of fine-grained structural changes, efficient error estimation is needed to accelerate the ALS flow~\cite{VECBEE22s,VACSEM24s,Dai25s}.
However, existing estimation methods evaluate LACs strictly at the \emph{primary outputs} (\emph{POs}) of the given netlist. 
They do not consider the impact when this approximated netlist is integrated into a larger design generated by HLS. 
To unlock the full potential of ALS within approximate HLS, an error estimation method is required to rapidly evaluate the impact of local LACs on the entire HLS-generated design.

\subsection{Error Metrics and Error Estimation Methods}\label{subsec:er_model}

Evaluating the accuracy of approximate circuits requires well-defined error metrics.
In this work, we focus on average error metrics, including \emph{mean absolute error} (\emph{MAE}), \emph{mean square error} (\emph{MSE}), \emph{mean absolute percentage error} (\emph{MAPE}), \emph{signal-to-noise ratio} (\emph{SNR}), and \emph{peak signal-to-noise ratio} (\emph{PSNR}).
Suppose a circuit has $\mathcal{T}$ input combinations and the probability of input combination $t$ is $P(t)$.
Let $\yexact(t)$ and $\yapprox(t)$ denote the exact and approximate outputs, respectively, for input combination $t$ and $\ymax$ be the maximum possible output value.
These error metrics are defined as follows:
\begin{equation}\label{eq:MAE}
\textit{MAE}=\sum^{\mathcal{T}}_{t=1}\left(P(t)\cdot\left|\yexact(t)-\yapprox(t)\right|\right),
\end{equation}
\begin{equation}\label{eq:MSE}
\textit{MSE}=\sum^{\mathcal{T}}_{t=1}\left(P(t)\cdot\left(\yexact(t)-\yapprox(t)\right)^2\right),
\end{equation}
\begin{equation}\label{eq:MAPE}
\textit{MAPE}=\sum^{\mathcal{T}}_{t=1}\left(P(t)\cdot\left|\frac{\yexact(t)-\yapprox(t)}{\yexact(t)}\right|\right),
\end{equation}
\begin{equation}\label{eq:SNR}
\textit{SNR}=10 \cdot \log_{10} \left( \frac{\sum_{t=1}^{\mathcal{T}} P(t) \cdot \yexact(t)^2}{\textit{MSE}} \right),
\end{equation}
\begin{equation}\label{eq:PSNR}
\textit{PSNR}=10 \cdot \log_{10} \left( \frac{\ymax^2}{\textit{MSE}} \right).
\end{equation}

In practice, the average error of an approximate circuit is typically evaluated using \emph{Monte Carlo} (\emph{MC}) \emph{simulation}~\cite{MonteCarlo16s}.
This method samples $T$ input combinations and calculates the average error across these samples as the estimated average error for the circuit.
However, MC simulation is slow for large designs.
To accelerate the evaluation process, researchers have proposed various error models to predict errors during approximate HLS.
They can be broadly categorized into two types: analytical models and \emph{machine learning} (\emph{ML})-based models~\cite{Marbell20s}.

Analytical models predict errors by propagating them through consecutive arithmetic instructions, achieving high accuracy in arithmetic-intensive datapaths~\cite{Li15s,Castro18s,Chen24s,Lee17s,AxHLS20s,PreDAC25s}.
However, practical HLS applications contain instructions that disrupt straightforward error propagation, such as memory accesses and branches.
The presence of these instructions leads to extensive case-by-case analysis in analytical models, limiting their generalizability~\cite{Lee17s}.

ML-based models, such as \emph{multi-layer perceptrons} (\emph{MLPs}) and \emph{graph neural networks} (\emph{GNNs}), have also been used to predict errors in approximate HLS~\cite{Zervakis19s,Sathidevi23s}.
These models can effectively capture complex error relationships even when memory-access and branch instructions are present.
However, ML-based models incur substantial inference latency per invocation, which limits their applicability in the iterative ALS step within the flow of approximate HLS.

Therefore, unlocking the power of ALS within HLS requires a fast error model that can handle complex situations when memory-access and branch instructions are present.

\section{Methodology}\label{sec:method}
We first formally define the optimization problem.

\textbf{Problem Formulation:} Given a high-level design description $C_{\textit{HLS}}$ representing the exact baseline hardware design $\mathcal{D}$, an input data distribution $\mathcal{P}_{\mathcal{T}}$, a user-specified error metric $\mathcal{E}$, and a corresponding error bound $e_b$, the objective of HALS is to generate an approximate hardware design $\mathcal{D}'$ that minimizes the total hardware cost while strictly satisfying $e_{\mathcal{D}'} \leq e_b$, where $e_{\mathcal{D}'}$ is the error of $\mathcal{D}'$ evaluated under the error metric $\mathcal{E}$ and the input distribution $\mathcal{P}_{\mathcal{T}}$.

\begin{figure}
\centering
\includegraphics[width=\linewidth]{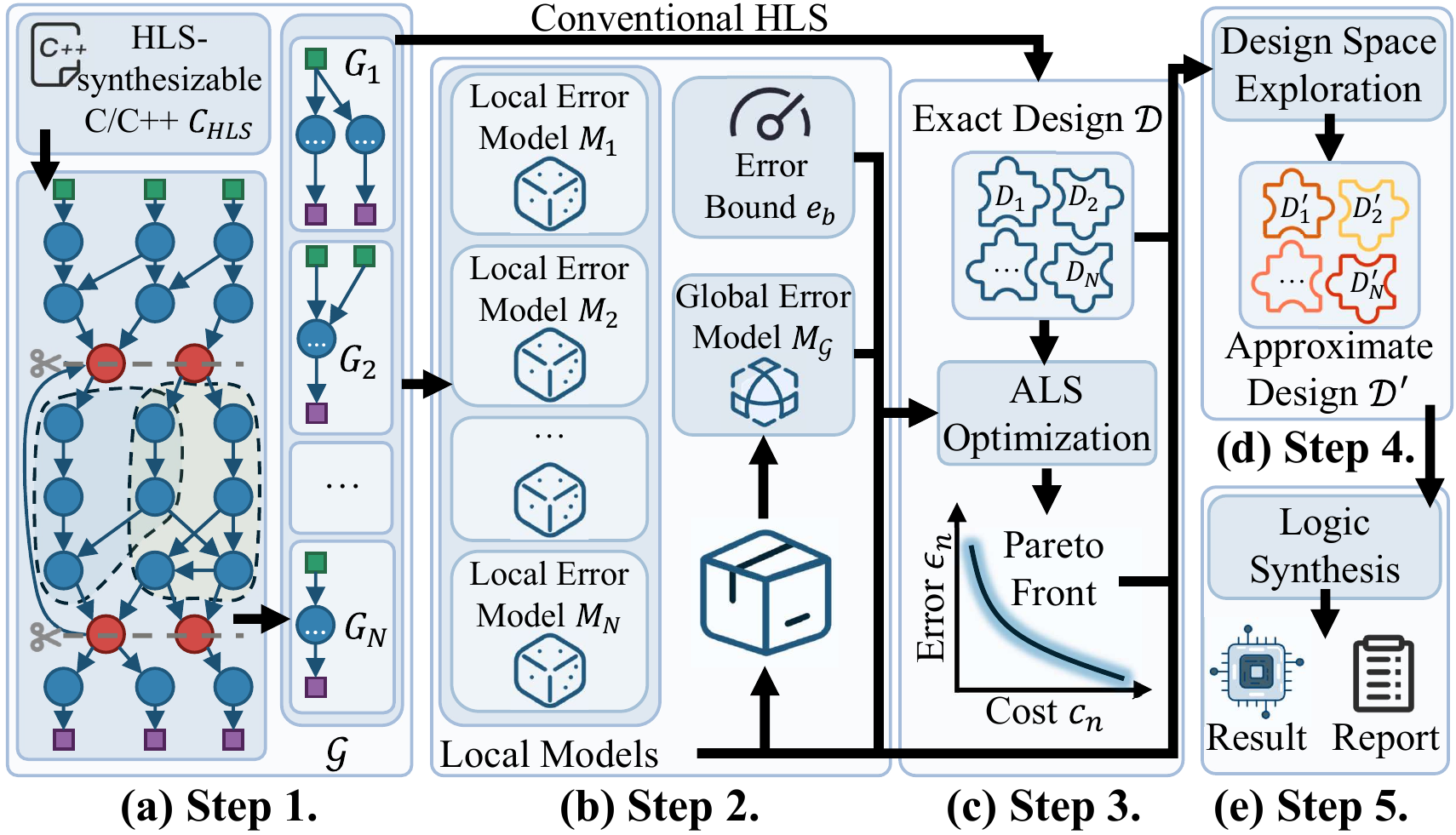}
\vspace{-0.8cm}
\caption{Overview of HALS: 
(a) Step~1: Convert the given C/C++ into a CDFG and partition it into ALS-manageable sub-graphs $\mathcal{G}=\left\{G_1,\dots,G_N\right\}$;
(b) Step~2: Construct a local error model $M_n$ for each sub-graph $G_n$ and a global error model $M_\mathcal{G}$ evaluating the combined impact of all sub-graphs;
(c) Step~3: Apply ALS to the exact design of each sub-graph;
(d) Step~4: Selectively replace sub-graphs with their approximate candidates through DSE;
(e) Step~5:Optimize the final design by logic synthesis and generate the corresponding report.
}
\vspace{-0.6cm}
\label{fig:flow}
\Description{Overview of HALS.}
\end{figure}

To solve this problem, HALS employs a five-step flow, as shown in Fig.~\ref{fig:flow}.
It takes $C_{\textit{HLS}}$, $\mathcal{P}_{\mathcal{T}}$, $\mathcal{E}$, and $e_b$ as inputs and proceeds as follows.
In Step~1 (Section~\ref{subsec:method-partition}), $C_{\textit{HLS}}$ is compiled into a CDFG, which is then partitioned into a set of ALS-manageable sub-graphs, $\mathcal{G}=\left\{G_1,\dots,G_N\right\}$.
Each $G_n\in\mathcal{G}$ corresponds to an exact hardware module $D_n$ generated by a conventional HLS flow, which collectively form the exact input design $\mathcal{D}$.
In Step~2 (Section~\ref{subsec:method-error}), a local error model $M_n$ is built for each sub-graph $G_n$, and a global error model $M_\mathcal{G}$ is constructed to characterize the combined impact of all sub-graphs on the final design error.
In Step~3 (Section~\ref{subsec:method-als}), guided by the local error model $M_n$, ALS generates a set of approximate candidates for each sub-graph $G_n$ based on $D_n$.
In Step~4 (Section~\ref{subsec:method-dse}), guided by the global error model $M_\mathcal{G}$, HALS selectively replaces sub-graphs in $\mathcal{G}$ with their approximate candidates.
This process assembles the approximate design $\mathcal{D}'$ that minimizes hardware cost while strictly ensuring its overall error $e_{\mathcal{D}'}$ is below $e_b$.
In Step~5 (Section~\ref{subsec:method-merge}), the assembled approximate design $\mathcal{D}'$ undergoes conventional logic synthesis and optimization to produce the final design.
The following subsections detail these steps.

\subsection{Step 1: Graph Partitioning}\label{subsec:method-partition}

In this step, the input high-level design description $C_{\textit{HLS}}$ is compiled and partitioned into a set of $N$ ALS-manageable sub-graphs $\mathcal{G}=\left\{G_1,G_2,\dots,G_N\right\}$, as shown in Fig.~\ref{fig:flow}(a).
This step begins by compiling $C_{\textit{HLS}}$ into an \emph{LLVM intermediate representation} (\emph{LLVM IR}) using \emph{Clang}, a high-performance and open-source compiler frontend~\cite{LLVM04s}.
During compilation, standard optimization techniques such as loop unrolling, function inlining, and abstract syntax tree flattening are applied to maximize the throughput of the design~\cite{unroll99s,fAST19s,inline22s}.
By analyzing the control and data dependencies among the IR instructions, a CDFG is constructed.
HALS then applies a two-stage partitioning method to this CDFG to divide it into ALS-manageable sub-graphs, as detailed below.


\subsubsection{Initial Partitioning}

Most ALS methods are only applicable to circuits represented as \emph{directed acyclic graphs} (\emph{DAGs}) and lack support for memory operations.
However, a CDFG generated by HLS may contain cycles and memory-access instructions.
Thus, ALS cannot be directly applied to HLS-generated designs.

In a CDFG, cyclic dependencies are induced by the back edges of branch instructions.
To guarantee that every resulting sub-graph is a DAG, HALS treats branch instructions as partition boundaries.
Furthermore, to ensure the partitioned sub-graphs are fully compatible with ALS, memory-access instructions are also used as partition boundaries.
As shown in Fig.~\ref{fig:partition}(b), gray dashed lines denote the cuts made at these boundaries, separating the memory-access and branch instructions from the other instructions.

To ensure the structural completeness of each obtained sub-graph, input and output ports are inserted at the boundaries.
Under this partitioning strategy, a single sub-graph may contain multiple input and output ports, as shown in Fig.~\ref{fig:partition}(c).

\subsubsection{Secondary Partitioning}

Conventional ALS is applied to the gate-level netlist, where the computational complexity of some steps grows exponentially with the gate count~\cite{BLASYS18s}.
For a large sub-graph from the initial partitioning, the gate count of the synthesized netlist can be excessive, leading to long ALS runtime.
Thus, it is necessary to further partition a large sub-graph into smaller ones while minimizing logic duplication across these smaller sub-graphs.
Although conventional partitioning methods (such as ratio cut) can reduce sub-graph size, they often introduce complex inter-partition dependencies that would complicate the subsequent error model~\cite{RatioCut89s}.
To avoid the complex dependencies while minimizing logic duplication, HALS adopts an \textit{output similarity-based clustering} method to perform the secondary partitioning.

Specifically, given a large sub-graph from the initial partitioning, HALS first groups its outputs into several clusters based on their logical similarities.
For each cluster, a new sub-graph is generated by retaining its outputs and their corresponding ancestor instructions.
This method avoids the complex inter-partition dependencies introduced by arbitrary graph cuts.
Meanwhile, clustering outputs with high similarity maximizes logic reuse within each sub-graph and minimizes logic duplication across sub-graphs.
For each obtained sub-graph, it is then simplified using standard LLVM passes.
Figs.~\ref{fig:partition}(c) and (d) show an example of this secondary partitioning process and the corresponding result, respectively.

Let $G_p$ be a sub-graph from the initial partitioning.
Assume it contains $k_p$ outputs $\left\{o_{p,1}, \ldots, o_{p,k_p}\right\}$ and $\theta_p$ IR instructions.
The middle sub-graph in Fig.~\ref{fig:partition}(c) shows an example with $k_p=3$ and $\theta_p=9$.
For each output $o_{p,i}$, a backward traversal collects all its ancestor instructions within $G_p$, forming an ancestor set $S_{p,i}$.



HALS uses the \textit{spectral clustering} method~\cite{SpectralClustering14s} to realize the output similarity-based clustering.
Specifically, it constructs a similarity matrix $\mathbf{A}_p\in [0,1]^{k_p \times k_p}$, where $\mathbf{A}_p[i,j]$ quantifies the similarity between outputs $o_{p,i}$ and $o_{p,j}$ using the Jaccard index of their ancestor sets~\cite{SpectralClustering14s}:
\begin{equation}
\mathbf{A}_p[i,j]=\text{Jaccard}\left(S_{p,i},S_{p,j}\right)=\frac{\left|S_{p,i}\bigcap S_{p,j}\right|}{\left|S_{p,i}\bigcup S_{p,j}\right|}.
\end{equation}
A higher $\mathbf{A}_p[i,j]$ value indicates a larger overlap in the ancestor instructions of $o_{p,i}$ and $o_{p,j}$.
The $i$-th row of $\mathbf{A}_p$ serves as a $k_p$-dimensional feature vector for $o_{p,i}$, projecting $o_{p,i}$ to a space based on its similarity to all other outputs.
HALS then applies $k$-means clustering to these row vectors to cluster the sub-graph outputs.

Although the similarity-based clustering method minimizes logic duplication across partitions, some logic duplication across different sub-graphs is inevitable because outputs in separate clusters may still share common ancestors.
For example, the dashed regions in Fig.~\ref{fig:partition}(c) outline two sub-graphs, $G_2$ and $G_3$, corresponding to two clusters.
Between $G_2$ and $G_3$, two overlapping ancestor instructions and one input are duplicated, as marked by an ``X'' in Fig.~\ref{fig:partition}(d).
While this redundancy might temporarily counteract hardware cost reduction, much of the duplicated logic is eliminated during the circuit optimization phase (detailed in Section~\ref{subsec:method-merge}).

\subsection{Step~2: Error Model Construction}\label{subsec:method-error}

In this step, we construct error models tailored for HALS, as shown in Fig.~\ref{fig:flow}(b).
As discussed in Sections~\ref{subsec:ALS} and \ref{subsec:er_model}, HALS requires a fast error estimation method that maps the error impact of sub-graph LACs to the entire HLS-generated design $\mathcal{D}$, while capable of handling memory-access and branch instructions.
To satisfy these requirements, we propose a scalable \textit{two-level regression model}, consisting of local error models and a global error model.

\subsubsection{Local Error Model}
For each sub-graph $G_n$, a local error model $M_n$ is established to guide the ALS.
Conventional ALS methods select candidate LACs based on their errors at the POs of the given netlist. 
In the context of HALS, this implies that the ALS methods can only evaluate the local errors at the outputs $\{o_{n,1},o_{n,2},\dots,o_{n,k_n}\}$ of the sub-graph $G_n$. 
We denote the error of $o_{n,i}$ as $\epsilon_{n,i}$.
When $G_n$ is approximated while the rest of the design remains exact, these local errors propagate, causing the entire design $\mathcal{D}$ to exhibit a system-level error $\epsilon_{n}$. 
To capture this relationship, $M_n$ is designed to map the local errors $\{\epsilon_{n,1},\epsilon_{n,2},\dots,\epsilon_{n,k_n}\}$ of $G_n$ to its induced system-level error $\epsilon_n$.

\begin{figure}
\centering
\begin{subfigure}[b]{0.49\linewidth}
\includegraphics[width=\linewidth]{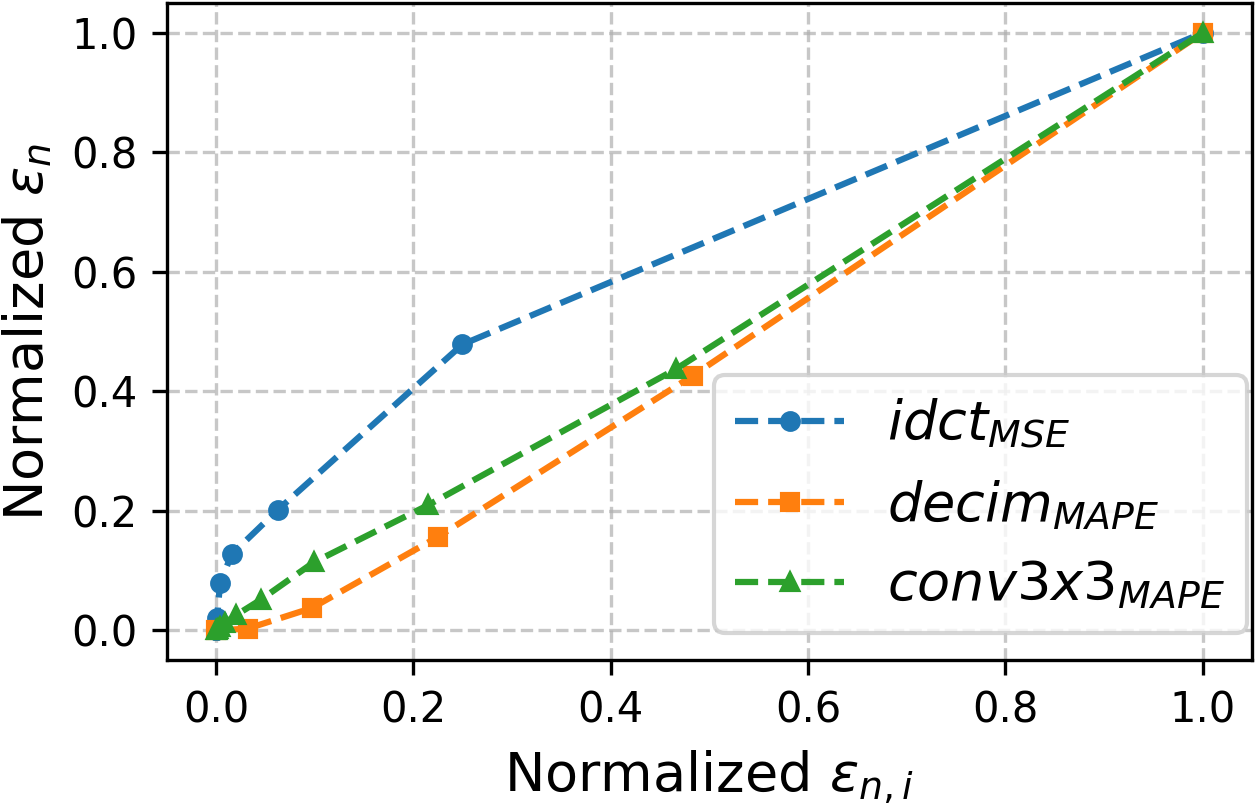}
\caption{}
\label{subfig:monotonicity}
\end{subfigure}
\begin{subfigure}[b]{0.49\linewidth}
\includegraphics[width=\linewidth]{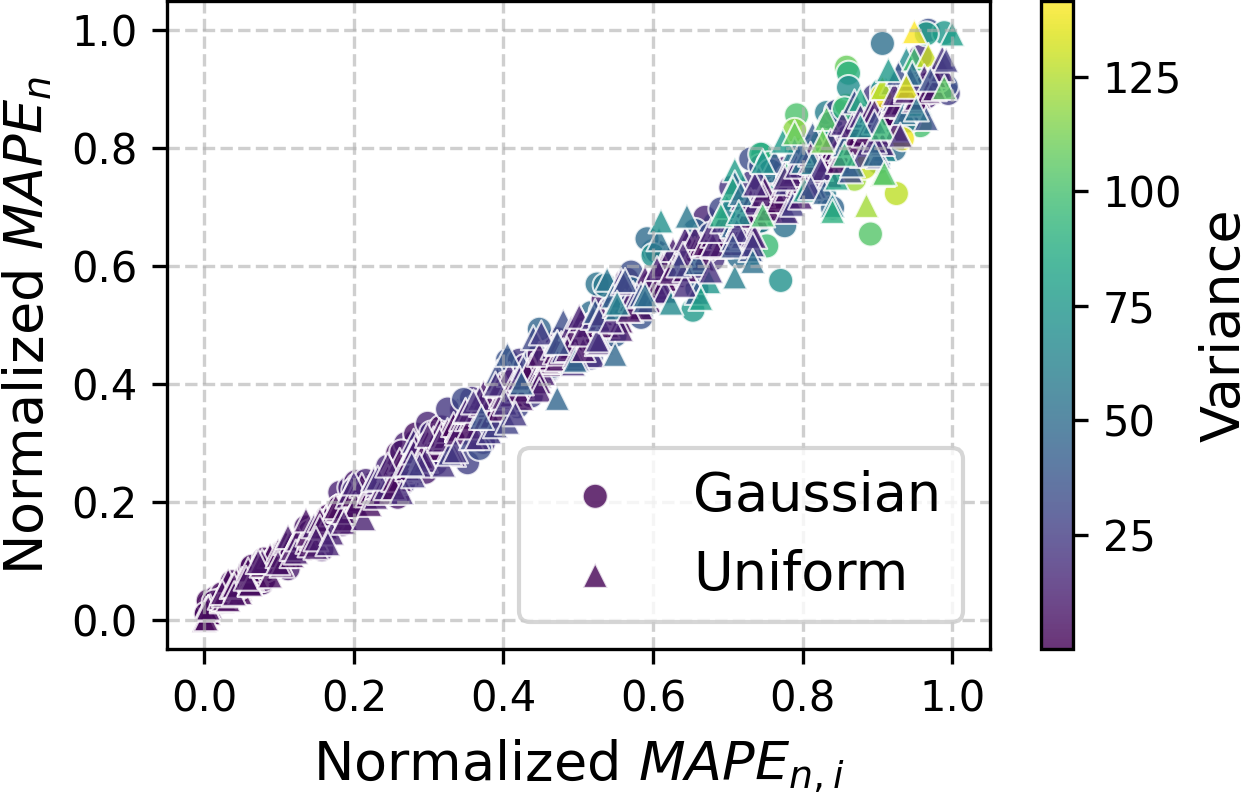}
\caption{}
\label{subfig:variance}
\end{subfigure}
\vspace{-0.4cm}
\caption{Experimental validation of error propagation properties.
(\subref{subfig:monotonicity}) Monotonic relationship between the normalized local error $\epsilon_{n,i}$ and the sub-graph-induced system-level error $\epsilon_n$ across different designs and error metrics.
(\subref{subfig:variance}) Distribution invariance observed in a 5-stage decimation filter, where Gaussian and uniform error distributions with different variances are injected into the third stage.
Despite different variances (indicated by color gradients), similar normalized $\text{MAPE}_{n,i}$ values result in comparable normalized $\text{MAPE}_n$.
}
\vspace{-0.4cm}
\Description{Experimental validation of error propagation properties.}
\label{fig:error-prelimiary}
\end{figure}

We first study the mapping properties by conducting experiments on some designs and error metrics.
Specifically, we conducted MC simulations by injecting errors into one output $o_{n,i}$ of a sub-graph, while maintaining the other outputs exact, and observed the corresponding system-level error $\epsilon_{n}$.
When the error magnitude is not excessive (quantified by $\text{MAPE} \le 30\%$ for each $o_{n,i}$), the average errors exhibit the following properties:
\begin{itemize}
\item \textit{Monotonicity}: For a given sub-graph output $o_{n,i}$, there exists a monotonic relationship between $\epsilon_{n,i}$ and $\epsilon_{n}$.
As shown in Fig.~\ref{fig:error-prelimiary}(\subref{subfig:monotonicity}), this monotonic behavior is consistently observed across different designs and error metrics.
\item \textit{Distribution invariance}: If two different approximation methods with different error distributions produce a similar average error $\epsilon_{n,i}$, they result in a comparable $\epsilon_{n}$.
Fig.~\ref{fig:error-prelimiary}(\subref{subfig:variance}) shows this using a 5-stage decimation filter.
Despite differences in the injected error variance (represented by color gradients), the relationship between the normalized $\text{MAPE}_{n,i}$ and $\text{MAPE}_{n}$ remains highly consistent.
\end{itemize}

Based on these observations, $M_n$ is formulated as follows:
\begin{equation}\label{eq:m_n}
M_n\coloneq \epsilon_{n} = \sum_{i=1}^{k_n} \alpha_{n,i} \cdot \left(\epsilon_{n,i}\right)^{\beta_{n,i}} + \gamma_{n},
\end{equation}
where $k_n$ is the number of outputs for $G_n$, and $\alpha_{n,i}$, $\beta_{n,i}$, and $\gamma_{n}$ are parameters to be fitted.
The model is considered reliable if $\gamma_{n}$ is relatively small and the coefficient of determination $R^2_{n}$ exceeds $0.85$; otherwise, a neural network is recommended as a fallback.



\subsubsection{Global Error Model}
Because each local error model $M_n$ evaluates an individual sub-graph, it ignores the error interactions with other sub-graphs. 
To capture these interactions for the subsequent DSE step (detailed in Section~\ref{subsec:method-dse}), a global model $M_{\mathcal{G}}$ is introduced.
Let $\epsilon_{\mathcal{G}}$ denote the \emph{concurrent global error} of design $\mathcal{D}$ when all sub-graphs are approximated simultaneously.
$M_{\mathcal{G}}$ is designed to map the individual sub-graph-induced system-level errors $\epsilon_{n}$ to this concurrent global error $\epsilon_{\mathcal{G}}$.
Similar to the local model, $M_\mathcal{G}$ is formulated as follows:
\begin{equation}\label{eq:m_g}
M_\mathcal{G}\coloneq \epsilon_{\mathcal{G}} = \sum_{n=1}^{N} \alpha_{n} \cdot \left(\epsilon_{n}\right)^{\beta_{n}} + \gamma,
\end{equation}
where $\alpha_{n}$, $\beta_{n}$, and $\gamma$ are parameters to be fitted.
The model is considered reliable if $\gamma$ is relatively small and $R^2_{\mathcal{G}}$ exceeds $0.85$.


\subsubsection{Model Fitting via IR-Level Simulation}
To fit $M_n$ and $M_{\mathcal{G}}$, HALS collects paired observations of $(\{\epsilon_{n,1},\epsilon_{n,2},\dots,\epsilon_{n,k_n}\}, \epsilon_n)$ and $(\{\epsilon_{1},\epsilon_{2},\dots,\epsilon_{N}\}, \epsilon_{\mathcal{G}})$ using LLVM IR-level MC simulations, which execute much faster than RTL or gate-level simulations. 
By executing these simulations across the entire design, HALS takes all memory-access and branch instructions into account and captures how local approximations aggregate and propagate through these instructions to the design outputs.


In each simulation round, input stimuli are sampled from the user-provided distribution $\mathcal{P}_{\mathcal{T}}$. 
To simulate the effect of ALS optimizations, HALS injects an unbiased, uniformly distributed error into each output $o_{n,i}$ of $G_n$ by rounding its least significant $\tau_{n,i}$ bits.
The rounding bitwidth $\tau_{n,i}$ ranges from 1 to $\left({\psi}_{n,i}-\Delta_{\psi}\right)$, where ${\psi}_{n,i}$ is the original bitwidth of $o_{n,i}$, and $\Delta_{\psi}$ is a user-specified parameter.
Using these randomized rounding configurations, HALS executes two distinct simulation phases to collect the fitting data:
\begin{itemize}
\item \emph{Local model fitting}: 
HALS approximates the outputs of one sub-graph $G_n$ at a time, leaving the rest of the design exact. 
This step collects the local errors $\{\epsilon_{n,1},\epsilon_{n,2},\dots,\epsilon_{n,k_n}\}$ (measured at the outputs of $G_n$) and $\epsilon_{n}$ (measured at the outputs of $\mathcal{D}$) to fit the $(2k_n + 1)$ parameters of $M_n$.
\item \emph{Global model fitting}: 
HALS samples several approximate configurations per sub-graph from the local fitting phase and combines them into a number of global configurations. 
Each combination corresponds to a vector of sub-graph-induced system-level errors $\{\epsilon_{1}, \epsilon_{2}, \dots, \epsilon_{N}\}$ measured in the local fitting phase.
Furthermore, the combination is applied to all sub-graphs simultaneously to yield the concurrent global error $\epsilon_{\mathcal{G}}$.
The data pairs $(\{\epsilon_{1}, \dots, \epsilon_{N}\}, \epsilon_{\mathcal{G}})$ over all combinations are used to fit the $(N + 1)$ parameters of $M_{\mathcal{G}}$.
\end{itemize}

Since $M_n$ and $M_{\mathcal{G}}$ contain only $(2k_n + 1)$ and $(2N + 1)$ parameters, respectively, the required numbers of simulation rounds to fit $M_n$ and $M_{\mathcal{G}}$ scale linearly with the number of sub-graph outputs $k_n$ and the total number of sub-graphs $N$, respectively.

\subsection{Step~3: ALS for Sub-Graph}\label{subsec:method-als}
As shown in Fig.~\ref{fig:flow}(c), this step performs ALS on each sub-graph $G_n$ identified in Step~1.
ALS takes three inputs: an exact circuit, an error metric, and a corresponding error bound.
The exact circuit is set as the exact hardware module $D_n$ generated for $G_n$ in Step~1.
Instead of using a conventional error metric, the ALS step of HALS is guided by the sub-graph-induced system-level error $\epsilon_n$, which is evaluated by applying the local error model $M_n$ constructed in Step~2 on the local output errors $\{\epsilon_{n,1}, \dots,\epsilon_{n,k_n}\}$ of the ALS-generated design. 
The error bound is set as the given bound $e_b$.


Most ALS methods operate as iterative and incremental optimization processes. 
They start from the exact circuit and progressively minimize the hardware cost while satisfying the given error bound.
During this progress, many intermediate approximate circuits are generated.
Although these intermediate circuits consume more hardware resources than the final optimized circuit, they exhibit lower errors.
Prior work has proposed to use these intermediate approximate circuits as candidates for subsequent DSE~\cite{QUADOL26s}.
Inspired by this, for each sub-graph $G_n$, HALS collects these intermediate circuits during the ALS process and extracts a Pareto front representing the optimized trade-offs between its induced system-level error $\epsilon_n$ and the corresponding hardware cost $c_n$.
These circuits then serve as the candidates for the subsequent DSE in Step~4.

\subsection{Step~4: Optimized ALS Design Selection}\label{subsec:method-dse}

In this step, as shown in Fig.~\ref{fig:flow}(d), HALS selectively replaces the exact sub-graphs with their ALS-generated approximate candidates.
The objective is to minimize the total hardware cost (\textit{e.g.}, area, delay, or power) without violating the global error bound $e_b$.

For each sub-graph $G_n$, Step~3 has already generated a Pareto front representing the optimized trade-offs between its induced system-level error $\epsilon_n$ and the corresponding hardware cost $c_n$.
To determine an optimized combination of these candidates across the entire design, HALS employs a greedy iterative DSE algorithm inspired by the search method in PreDAC~\cite{PreDAC25s}.
The search is guided by a \emph{cost-error ratio}.
Specifically, assume that the current candidate for $G_n$ has a cost $c_n$ and an induced system-level error $\epsilon_n$.
If a new candidate yields a lower cost $c_{n'}$ but a higher induced error $\epsilon_{n'}$, the cost-error ratio is calculated as:
\begin{equation}
\text{cost-error ratio} = \frac{\Delta \text{cost}}{\Delta \epsilon_{\mathcal{G}}} = \frac{c_n - c_{n'}}{\alpha_n \cdot \left({\left(\epsilon_{n'}\right)}^{\beta_n} - {\left(\epsilon_n\right)}^{\beta_n}\right)},
\end{equation}
where the denominator is derived from the global error model $M_{\mathcal{G}}$.
A higher cost-error ratio indicates a more favorable trade-off between the error and the hardware cost.

At each iteration, HALS greedily selects the candidate with the maximum cost-error ratio over all sub-graphs to replace its corresponding current candidate.
The iterative replacement process continues until the global error predicted by $M_\mathcal{G}$ reaches the bound $e_b$, or all approximate candidates have been considered.
The obtained approximate design is then validated by a gate-level MC simulation.
If the actual simulated error exceeds $e_b$, the DSE process triggers a rollback mechanism: it undoes the most recent replacement, reverting to the preceding design, and re-validates it.
This rollback continues until an approximate design that strictly satisfies $e_b$ is found.
Then, the obtained approximate design $\mathcal{D}'$ is passed to Step~5 for final logic synthesis and optimization.

\subsection{Step~5: Circuit Optimization}\label{subsec:method-merge}
In this step, as shown in Fig.~\ref{fig:flow}(e), the assembled approximate design $\mathcal{D}'$ from Step~4 undergoes full-design logic synthesis and optimization to yield the final hardware implementation.
Synthesizing $\mathcal{D}'$ as a whole is essential for two reasons.
First, it allows conventional synthesis tools to globally identify and eliminate the temporary logic duplication introduced by the secondary partitioning in Step~1.
Second, it unlocks more optimization opportunities. 
Techniques such as retiming~\cite{Retiming06s} and register packing~\cite{RegPack04s} can exploit these opportunities to further improve area, delay, and power.
Finally, this step outputs the optimized approximate design alongside a report on circuit area, delay, power, and error.

\section{Experimental Results}\label{sec:exp}

\subsection{Experimental Setup}\label{subsec:exp-setup}
The HLS was implemented using an LLVM-based front-end~\cite{LLVM04s} and an in-house developed backend.
For ALS, we utilized a modified version of ResubALS~\cite{ResubALS25s} that integrates the local error models for the partitioned sub-graphs.
Logic synthesis and power analysis were performed using Yosys~\cite{YOSYSs} and Synopsys Design Compiler~\cite{DesignCompiler}, respectively, while Verilator~\cite{Verilator18s} was used for RTL simulation.
Consistent with prior works~\cite{ADVISOR25s,PreDAC25s}, the synthesis flow utilized the Nangate 45nm library~\cite{Nangate}.
All experiments were conducted on a server with an AMD Ryzen 9 5950X CPU at 3.4GHz using 16 threads and 32GB of DDR4 memory.

We evaluated HALS on 14 benchmarks. 
Specifically, 5 applications were sourced from S2CBench~\cite{S2CBench14s}, 4 from PaderBench~\cite{PaderBench19s}, and 5 from PolyBench~\cite{Polybench12s}.
Table~\ref{tab:case-list} details the characteristics of these benchmarks, including the number of simulation patterns used for error modeling (denoted as \textit{\#Sim. Patterns}), the target hardware optimization metric (\textit{Metric}), the baseline cost of the original exact circuit evaluated under the corresponding metric (\textit{Orig. Cost}), and the final number of partitioned sub-graphs (\textit{\#Sub-graphs}).

\begin{table}
\caption{Characteristics of the evaluated benchmarks.
The baseline cost is measured in ${\mu m}^2$ for area and $\mu W$ for power.}
\vspace{-0.4cm}
\small
\label{tab:case-list}
\tabcolsep=4pt
\begin{tabular}{lcccc}
\toprule
Benchmark & \#Sim. Patterns & Metric & Orig. Cost & \#Sub-graphs \\
\midrule
fir9~\cite{S2CBench14s}        & 10,000         & Area  & 465.13   & 1  \\
fft~\cite{S2CBench14s}         & 10,000         & Area  & 76281    & 16 \\
interp~\cite{S2CBench14s}      & 10,000         & Area  & 49932    & 4  \\
decimation~\cite{S2CBench14s}  & 10,000         & Area  & 12489    & 5  \\
Sobel~\cite{S2CBench14s}       & $512\times512$ & Area  & 556.56   & 1  \\
RGB2YCbCr~\cite{PaderBench19s} & 10,000         & Area  & 1673.7   & 1  \\
conv3x3~\cite{PaderBench19s}   & 10,000         & Power & 4950.0   & 1  \\
fir13~\cite{PaderBench19s}     & 10,000         & Area  & 1176.9   & 1  \\
Gauss.Blur~\cite{PaderBench19s}& $512\times512$ & Power & 428.86   & 1  \\
cholesky~\cite{Polybench12s}   & 65,536         & Area  & 37389    & 3  \\
gesummv~\cite{Polybench12s}    & 65,536         & Area  & 9370.1   & 2  \\
jacobi-1d~\cite{Polybench12s}  & 65,536         & Area  & 12669    & 2  \\
jacobi-2d~\cite{Polybench12s}  & 65,536         & Area  & 14702    & 2  \\
syr2k~\cite{Polybench12s}      & 65,536         & Area  & 7479.9   & 2  \\
\bottomrule
\end{tabular}
\vspace{-0.8cm}
\end{table}

To ensure realistic evaluation, the input distributions were tailored for each benchmark.
For \textit{fir9}, \textit{fir13}, \textit{interp}, and \textit{decimation}, the inputs comprise data points that are uniformly distributed and coefficients following specific mathematical distributions.
For \textit{Sobel} and \textit{Gauss.Blur}, inputs are derived from real-world image pixel distributions.
For all other benchmarks, including \textit{conv3x3}, \textit{fft}, \textit{RGB2YCbCr}, and the PolyBench benchmarks, all inputs are uniformly distributed.

To ensure fair comparisons and demonstrate the generality of HALS, the target hardware cost metrics were selected to match those of prior works~\cite{ADVISOR25s,PreDAC25s}.
As shown in Table~\ref{tab:case-list}, circuit area is evaluated for most benchmarks, whereas power consumption is only used for \textit{conv3x3} and \textit{Gauss.Blur}.
The design quality is reported using the \textit{hardware cost ratio}, defined as the ratio of the hardware cost of the approximate design over that of the original exact design.
A lower ratio indicates a better result.

In some small benchmarks, boundaries for the initial partitioning (\textit{e.g.}, branch and memory-access instructions) are either absent or localized near the inputs, resulting in a single sub-graph after the initial partitioning.
Furthermore, this sub-graph contains only one output.
Therefore, the secondary partitioning is bypassed, yielding a single indivisible sub-graph for these benchmarks, as reflected in Table~\ref{tab:case-list}.
Consequently, no error models were constructed for them.
Instead, the errors during the ALS and DSE steps were evaluated through direct MC simulations.
For the other benchmarks, the number of sub-graphs is determined by the partitioning step, where the number of centroids for the $k$-means clustering process in the secondary partitioning is adjustable.
More centroids yield more, but smaller, sub-graphs. 
Since the runtime of the ALS step grows exponentially with the sub-graph gate count, increasing the centroids reduces the overall HALS runtime. 
However, more sub-graphs also increase logic duplication, thereby degrading the design quality. 
On the other hand, fewer centroids produce larger sub-graphs that can exhaust server memory during the ALS step. 
Therefore, the number of centroids was empirically minimized to maximize design quality while strictly adhering to the server memory constraint.

\begin{table}
\centering
\tabcolsep=3pt
\small
\caption{Performance comparison between the proposed model and the MLP baseline.
For our method, the construction time is the total time it takes to construct all local models and the global model, while the inference time is reported separately for the local and global models.}
\vspace{-0.4cm}
\label{tab:error_model_comp}
\begin{tabular}{@{}l|cc|ccc|cc@{}}
\toprule
\multirow{2}{*}{Benchmark} & \multicolumn{2}{c|}{Construction Time (s)} & \multicolumn{3}{c|}{Inference Time ($\mu\text{s}$)} & \multicolumn{2}{c}{Accuracy ($R^2$)} \\
\cmidrule(l){2-8} 
& \multirow{2}{1cm}{\centering Ours (Total)} & \multirow{2}{*}{MLP} & \multicolumn{2}{c}{Ours} & \multirow{2}{*}{MLP} & \multirow{2}{1cm}{\centering Ours ($M_\mathcal{G}$)} & \multirow{2}{*}{MLP} \\
\cmidrule(lr){4-5}
& & & $M_n$ & $M_\mathcal{G}$ & & & \\
\midrule
fft        & 18.2 & 23.0 & 0.44 & 0.36 & 1.25   & \textbf{1.00}  & 0.98    \\
interp     & 0.82 & 4.35 & 0.04 & 0.04 & 1.21   & \textbf{1.00}  & \textbf{1.00}    \\
decimation & 3.50 & 1.54 & 0.11 & 0.09 & 1.17   & 0.91  & \textbf{0.94}    \\
cholesky   & 1.89 & 4.68 & 0.10 & 0.06 & 1.20   & \textbf{0.96}  & \textbf{0.96}    \\
gesummv    & 15.5 & 3.50 & 0.05 & 0.03 & 1.20   & \textbf{1.00}  & \textbf{1.00}    \\
jacobi-1d  & 96.1 & 37.4 & 0.04 & 0.04 & 1.11   & 0.98  & \textbf{0.99}    \\
jacobi-2d  & 153  & 58.7 & 0.04 & 0.03 & 1.11   & \textbf{1.00}  & \textbf{1.00}    \\
syr2k      & 0.42 & 1.11 & 0.03 & 0.03 & 1.34   & \textbf{1.00}  & \textbf{1.00}    \\
\midrule
Geomean    & 7.44 & 7.06 & 0.07 & 0.06 & 1.20   & \textbf{0.98}  & \textbf{0.98}    \\
\bottomrule
\end{tabular}
\vspace{-0.6cm}
\end{table}

\begin{table*}
\caption{Comparison of hardware cost ratio and runtime (\textit{RT}) for ADVISOR, PreDAC, and HALS under various error bounds $e_b$.
}
\vspace{-0.4cm}
\small
\label{tab:final-results}
\begin{tabular}{l|c|ccccc|ccccc}
\toprule
 \multirow{2}{*}{Benchmark}& \multirow{2}{1.2cm}{\centering Error Metric $\mathcal{E}$} & \multirow{2}{1.4cm}{\centering Tight Error Bound $e_b$} & \multicolumn{2}{c}{Prior Works} & \multicolumn{2}{c|}{HALS (Ours)} & \multirow{2}{1.4cm}{\centering Loose Error Bound $e_b$} & \multicolumn{2}{c}{Prior Works} & \multicolumn{2}{c}{HALS (Ours)} \\
 \cmidrule(lr){4-5}\cmidrule(lr){6-7}\cmidrule(lr){9-10}\cmidrule(lr){11-12}
& &  &  Cost Ratio  &  RT (s)   &  Cost Ratio  &  RT (s) &  &  Cost Ratio  &  RT (s) &  Cost Ratio  &  RT (s) \\
\midrule
fir9       & MAPE & 10\%      & 57.59\% & 840    & \textbf{50.03\%} & \textbf{18.50}  & 20\% & 39.27\% & 840    & \textbf{32.65\%} & \textbf{19.81}   \\
fft        & MAPE & 10\%      & 41.71\% & 1980   & 93.06\% & 4944.3 & 20\% & 28.71\% & 1980   & 89.81\% & 4945.4  \\
interp     & MAPE & 10\%      & 54.25\% & 3540   & \textbf{38.86\%} & 6439.8 & 20\% & 46.67\% & 3540   & \textbf{29.83\%} & 6440.1  \\
decimation & MAPE & 10\%      & 46.59\% & 5580   & 56.24\% & \textbf{3206.1} & 20\% & 34.26\% & 5580   & 49.11\% & \textbf{3208.0}  \\
Sobel      & PSNR & 20dB         & 39.35\% & 2580   & \textbf{34.80\%} & \textbf{77.97}  & 10dB    & 18.92\% & 2580   & 22.74\% & \textbf{86.91}   \\
\midrule
RGB2YCbCr  & MAE  & 1000         & 55.32\% & 1731   & \textbf{29.28\%} & \textbf{1455.6} & 1400    & 54.18\% & 1992   & \textbf{28.20\%} & \textbf{1460.1}  \\
conv3x3    & MAE  & 1000         & 38.40\% & 97.8   & \textbf{34.55\%} & 20112  & 1400    & 36.80\% & 147    & \textbf{29.29\%} & 23185 \\
fir13      & SNR  & 25dB         & 77.00\% & 14     & \textbf{17.44\%} & 193.96 & 19dB    & 56.00\% & 17     & \textbf{25.93\%} & 170.47  \\
Gauss.Blur & SNR  & 25dB         & 51.90\% & 218    & \textbf{32.58\%} & \textbf{32.28}  & 19dB    & 41.60\% & 274    & \textbf{14.52\%} & \textbf{36.59}   \\
\midrule
Geomean    &      &  & 50.21\% & 706.39 & \textbf{38.89\%}  & \textbf{669.42} &         & 37.82\% & 786.85 & \textbf{31.48\%}  & \textbf{693.58}  \\
\bottomrule
\end{tabular}
\vspace{-0.4cm}
\end{table*}

\subsection{Evaluation of the Error Model}\label{subsec:exp-error-model}
To evaluate the proposed error model, we compare its performance against a neural network baseline. 
Specifically, we trained a unified MLP to replace both the local models ($M_n$) and the global model ($M_{\mathcal{G}}$). 
The MLP consists of two hidden layers with 64 and 32 neurons, respectively, and ReLU activations.
It is trained on the same MC simulation datasets with the parameter $\Delta_\psi$ fixed at 3.
Because the MLP serves as a unified predictor, its input layer is designed to accept the errors from all sub-graph outputs across the entire design. 
In the ALS step for a specific sub-graph $G_n$, we activate the input neurons corresponding to the errors of $G_n$'s outputs ($\{\epsilon_{n,1}, \dots, \epsilon_{n,k_n}\}$), and set all other inputs to zero.
This allows the MLP to mimic $M_n$ and predict the sub-graph-induced system-level error $\epsilon_n$.
In the DSE step, the MLP evaluates the combined errors from all candidate sub-graphs concurrently, substituting $M_{\mathcal{G}}$ to predict the global error $\epsilon_{\mathcal{G}}$.

Table~\ref{tab:error_model_comp} details the construction time, inference time, and final accuracy for the MLP-based error model and our proposed model on benchmarks with more than one sub-graph in Table~\ref{tab:case-list}. 
The single-sub-graph benchmarks are not considered as their errors are evaluated via direct MC simulations without using the error models.
The accuracy is quantified by $R^2$, with the highest value for each benchmark highlighted in bold.
For our model, the construction time includes dataset generation and fitting for all local models and the global model.
For the baseline MLP model, it includes global dataset generation and MLP training time.
The results indicate that both methods demand comparable construction time and achieve nearly identical prediction accuracy with average $R^2 \approx 0.98$.
However, our method shows an advantage in inference efficiency.
Because our models avoid the matrix multiplications in MLPs, our local models ($M_n$) evaluate $17.1\times$ faster on average than the MLP in the ALS step.
Similarly, our global model ($M_{\mathcal{G}}$) achieves a $20\times$ speedup over the MLP in the DSE step.
\subsection{Comparison with State-of-the-Art Methods}
To evaluate HALS against state-of-the-art approximate HLS methods, we compare it with ADVISOR~\cite{ADVISOR25s} and PreDAC~\cite{PreDAC25s}. 
Table~\ref{tab:final-results} shows the hardware cost ratios and runtimes (denoted as \textit{RT}) across various error bounds $e_b$.
The results for the compared methods are taken from their papers. 
For a fair comparison, our evaluation focuses on the 9 benchmarks from S2CBench and PaderBench used in their respective studies. 
Specifically, the S2CBench benchmarks (the first five in Table~\ref{tab:final-results}) are compared against ADVISOR, whereas the PaderBench benchmarks (the subsequent four) are compared against PreDAC. 
Both baselines are collectively denoted as ``Prior Works'' in Table~\ref{tab:final-results}. 
Note that due to differences in computational platforms, the reported runtimes are provided for reference only.

As shown in Table~\ref{tab:final-results}, HALS achieves lower hardware costs under the same error constraints for most benchmarks. 
Overall, HALS reduces the average hardware cost ratio by 11.32\% compared to the baselines under tight error constraints, and by 6.34\% under loose constraints. 
Two exceptions are \textit{decimation} and \textit{fft}, which we analyze in detail below.

\textit{Decimation} is a pipelined design dominated by multiply-\linebreak accumulate instructions across all stages.
This architecture causes rapid error amplification as approximations in early stages propagate downstream.
Therefore, HALS must enforce strict local error bounds on these early stages, meaning that most hardware cost reduction is derived from the final stages of the pipeline.

The highly connected butterfly structure of \textit{fft} causes outputs to share a large number of ancestor instructions. 
Our secondary partitioning duplicates many shared ancestors across sub-graphs, which resulted in a $3.19\times$ intermediate area increase before the ALS step. 
This massive duplication cannot be fully eliminated during the circuit optimization in Step~5. 
To address highly connected CDFGs like \textit{fft}, we plan to introduce a duplication factor in the future to characterize sub-graph topology. 
A predefined threshold for this factor will allow HALS to dynamically bypass secondary partitioning, preventing excessive logic duplication.

Regarding the runtime, the ALS step dominates the overall runtime of HALS. 
Because the runtime complexity of certain steps in ALS grows exponentially with the gate count, the runtime is highly sensitive to the size of individual sub-graphs rather than the total circuit area. 
For example, although \textit{conv3x3} does not possess the largest circuit area among the benchmarks, it cannot be partitioned, resulting in the longest runtime. 
In contrast, while \textit{fft} exhibits the largest circuit area, it is partitioned into 16 independent sub-graphs. 
Because the ALS for these sub-graphs can be executed in parallel, its overall runtime is shorter than \textit{conv3x3}.

\begin{figure}[htbp]
\centering
\begin{subfigure}[b]{0.49\linewidth}
\includegraphics[width=\linewidth]{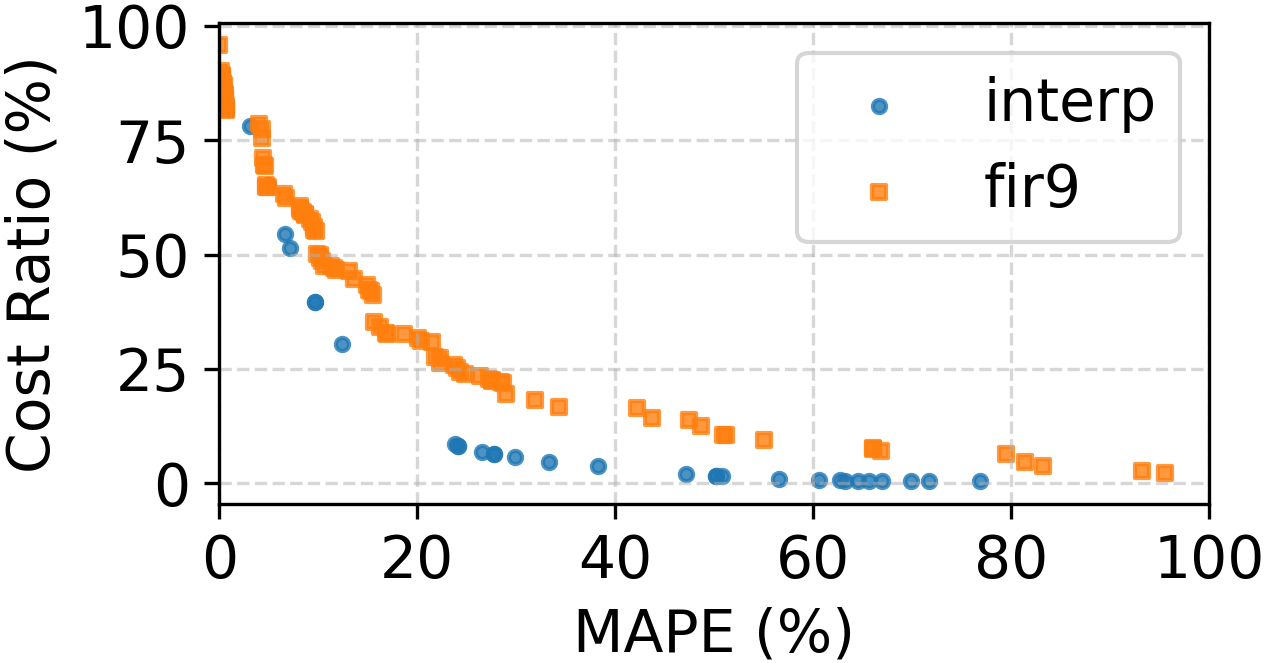}
\caption{}
\label{subfig:exp-fir9-interp}
\end{subfigure}
\begin{subfigure}[b]{0.49\linewidth}
\includegraphics[width=\linewidth]{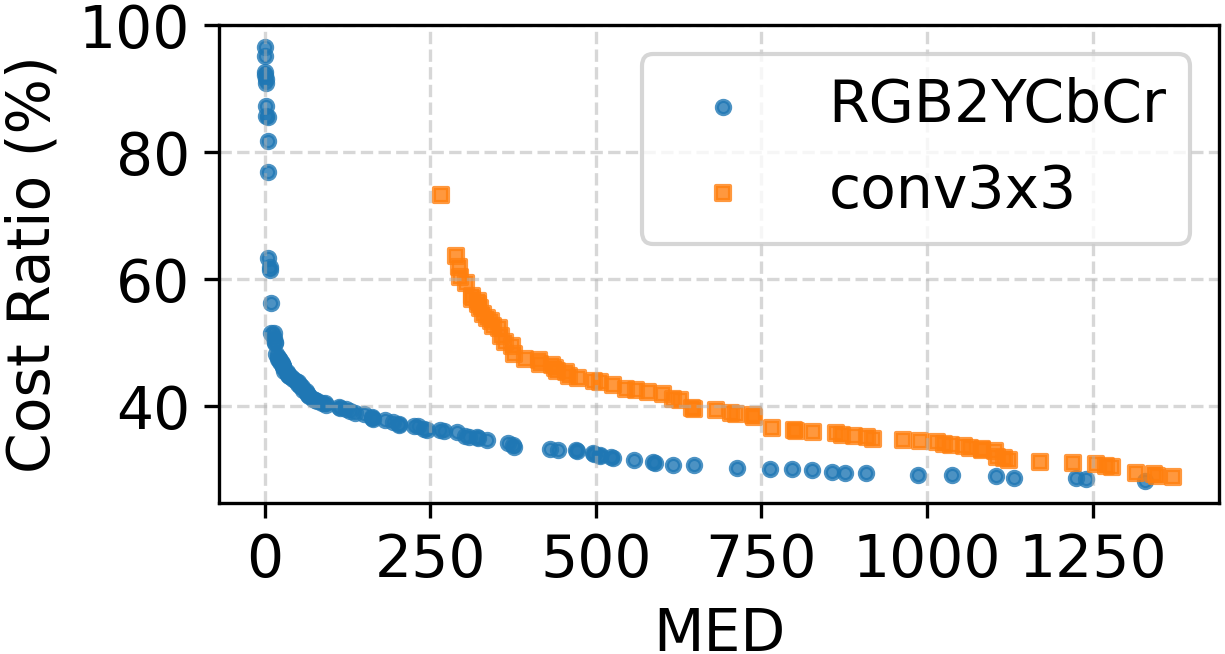}
\caption{}
\label{subfig:exp-conv3-rgb}
\end{subfigure}
\vspace{-0.4cm}
\caption{Pareto fronts of hardware cost ratio vs. error for $4$ benchmarks optimized by HALS.}
\vspace{-0.6cm}
\Description{Pareto fronts of hardware cost vs. errors for HALS}
\label{fig:exp-pareto}
\end{figure}

To further evaluate HALS, we plot the Pareto fronts for 4 benchmarks across a broader range of error bounds in Fig.~\ref{fig:exp-pareto}. 
The results show that HALS achieves substantial hardware savings under any given error constraint.

\subsection{Scalability on Complex Designs}
While HALS shows superior performance on the benchmarks from S2CBench and PaderBench, most of them are relatively small and lack complex memory interactions. 
To evaluate the generality and scalability of HALS on large-scale and memory-intensive benchmarks, we extended our evaluation to five benchmarks from PolyBench~\cite{Polybench12s}. 
Table~\ref{tab:polybench} shows the results, including the runtime and area ratio, under a $10\%$ MAPE constraint. 
Across these benchmarks, HALS achieved an average area ratio of $54.22\%$. 
The results show the robustness and scalability of the proposed HALS method.

\begin{table}[htbp]
\vspace{-0.4cm}
\caption{Optimization results of HALS for PolyBench benchmarks under a $10\%$ MAPE error bound.}
\small
\label{tab:polybench}
\centering
\vspace{-0.4cm}
\begin{tabular}{lccc}
\toprule
Benchmark & Orig. Area ($\mu m^2$) & Runtime (s) & Area Ratio \\
\midrule
cholesky    & 37389 & 553.04 & 67.46\% \\
gesummv     & 9370.1  & 250.21 & 72.34\% \\
jacobi-1d   & 12669 & 157.48 & 44.52\% \\
jacobi-2d   & 14702 & 147.68 & 33.77\% \\
syr2k       & 7479.9  & 393.93 & 63.87\% \\
\midrule
Geomean     & 13731   & 263.39 & 54.22\% \\
\bottomrule
\end{tabular}
\vspace{-0.6cm}
\end{table}

\section{Conclusion}\label{sec:conclude}

This paper introduces HALS, a framework that integrates ALS into the flow of approximate HLS.
By incorporating fine-grained approximation induced by ALS, HALS expands the design space of approximate HLS and enables the generation of higher-quality approximate circuits.
Experimental results show that under the same error bound, HALS reduces the average hardware cost by 11\% compared to existing state-of-the-art methods.
In the future, we plan to optimize the graph partitioning strategy to further reduce the hardware overhead introduced by partitioning.

\bibliographystyle{unsrt} 
\bibliography{refs}

\end{document}